\newif\ifosa
\osafalse       
\ifosa
    \documentclass{osa-article}
    \journal{oe}
\else
    \documentclass{article}
    \usepackage{fullpage}
    \usepackage[affil-it]{authblk}
    \usepackage{amsmath}
    \usepackage{amssymb}
    \usepackage{cite}
\fi

\usepackage{tikz}
\usetikzlibrary{arrows.meta,decorations.markings}
\newcommand{\M}[1]{\mathbf{#1}}
\newcommand{\T}[1]{\mathrm{#1}}
\newcommand{\V}[1]{\boldsymbol{#1}}

\renewcommand{\Im}{\M{I}}

\newcommand{\Rm}{\M{R}}
\newcommand{\Xm}{\M{X}}

\newcommand{\Vm}{\M{V}}

\newcommand{\subto}{\T{s.t.}}

\newcommand{\approxkazero}{\stackrel{ka\ll1}{\approx}}

\newcommand{\myshift}{\mkern-4.5mu}
\providecommand*{\unit}[1]{\ensuremath{\mathrm{\,#1}}}

\definecolor{red}{RGB}{228,26,28}
\definecolor{blue}{RGB}{55,126,184}
\definecolor{green}{RGB}{77,175,74}

\begin{document}

\ifosa
\else
    \title{Trade-offs in absorption and scattering by nanophotonic structures}
    
    \author{K. Schab\thanks{Department of Electrical and Computer Engineering Santa Clara University, Santa Clara, CA, USA}, A. Rothschild$^*$, K. Nguyen$^*$, M. Capek\thanks{Department of Electromagnetic Field, Czech Technical University in Prague, Prague, Czech Republic}, L. Jelinek$^\dagger$, and M. Gustafsson\thanks{Department of Electrical and Information Technology, Lund University, Lund, Sweden}}
	\date{}

    \maketitle
\fi



\begin{abstract}
Trade-offs between feasible absorption and scattering cross sections of obstacles confined to an arbitrarily shaped volume are formulated as a multi-objective optimization problem solvable by Lagrangian-dual methods.  Solutions to this optimization problem yield a Pareto-optimal set, the shape of which reveals the feasibility of achieving simultaneously extremal absorption and scattering.  Two forms of the trade-off problems are considered involving both loss and reactive material parameters.  Numerical comparisons between the derived multi-objective bounds and several classes of realized structures are made.  Additionally, low-frequency (electrically small, long wavelength) limits are examined for certain special cases.
\end{abstract}


\section{Introduction}
Recently, significant effort has been directed toward bounding parameters related to the performance of plasmonic and photonic devices, e.g., absorption and scattering cross sections of metallic particles at optical frequencies.   These bounds quantify the feasibility of design objectives and provide concrete references by which to assess inverse design routines aimed at automatically synthesizing devices meeting prescribed performance criteria \cite{molesky2018inverse,angeris2019computational,molesky2020hierarchical}.  To this end, bounds on single-frequency nanophotonic performance have been developed at varying levels of design constraint specificity, with more general bounds covering broader scenarios and more specific methods providing tighter physical limits for stricter design constraints.  This categorization is not binary, and previous methods of deriving such bounds increase in specificity from shape-independent bounds~\cite{miller2016fundamental}, shape-dependent bounds~\cite{molesky2020fundamental}, and finally, shape- and material-dependent bounds~\cite{gustafsson2020upper,trivedi2020fundamental,kuang2020maximal,molesky2020t,molesky2020global}.  Broadband performance bounds have also been established using sum rules \cite{sohl2007physical} and assumptions regarding specific response profiles \cite{shim2020optical}.

One approach in the development of single-frequency shape- and material-dependent physical limits is to maximize or minimize a particular quantity (e.g., absorption cross section) by forming duality-based convex optimization problems over polarization currents confined to a prescribed region and subject to physically-motivated constraints, e.g., conservation of power \cite{gustafsson2020upper}. The convexity of the underlying optimization problems guarantees that their unique solutions are global extrema which are found with convex optimization tools~\cite{boyd2004convex}, i.e., no heuristic is needed.  By nature of the formulation of these optimization problems, the derived bounds implicitly apply to all substructures comprised of the design material within the design region, regardless of geometric complexity.  Though the form of objectives and constraints vary from problem to problem, the general strategy closely resembles that used to derive broad classes of physical limitations on the efficiency, directivity, and bandwidth of microwave antennas \cite{uzsoky1956theory,gustafsson2016antenna,2018_Shahpari_TAP,capek2017minimization,gustafsson2019tradeoff}.

This work aims to study the coupled nature of absorption and scattering, answering the question: \emph{Are the best absorbers also the best scatterers?} To carry this out, we adapt single-objective formulations for bounds on absorption and scattering cross sections into multiobjective forms~\cite{1978CohonMultiobjectiveProgrammingAndPlanning,
Deb_MultiOOusingEA,boyd2004convex} and study the resulting feasible objective space.  In particular, this work examines the relative trade-off between optimal absorption and scattering in cases where both measures are maximized as well as cases where one metric is maximized and the other is minimized, e.g., maximal absorption with minimum scattering.

This paper is structured as follows. Multi-objective optimization problems involving scattering and absorption are introduced in Sec.~\ref{sec:MOO}, accompanied by definitions of necessary quantities and their physical meaning. The particular solution of optimal trade-offs for prescribed material losses and for fully prescribed materials are examined in Secs.~\ref{sec:losses} and~\ref{sec:materials}, respectively. The paper is concluded in Sec.~\ref{sec:conclusions}.


\section{Multiobjective bounds}
\label{sec:MOO}

\begin{figure}
    \centering
    \includegraphics[width=3.5in]{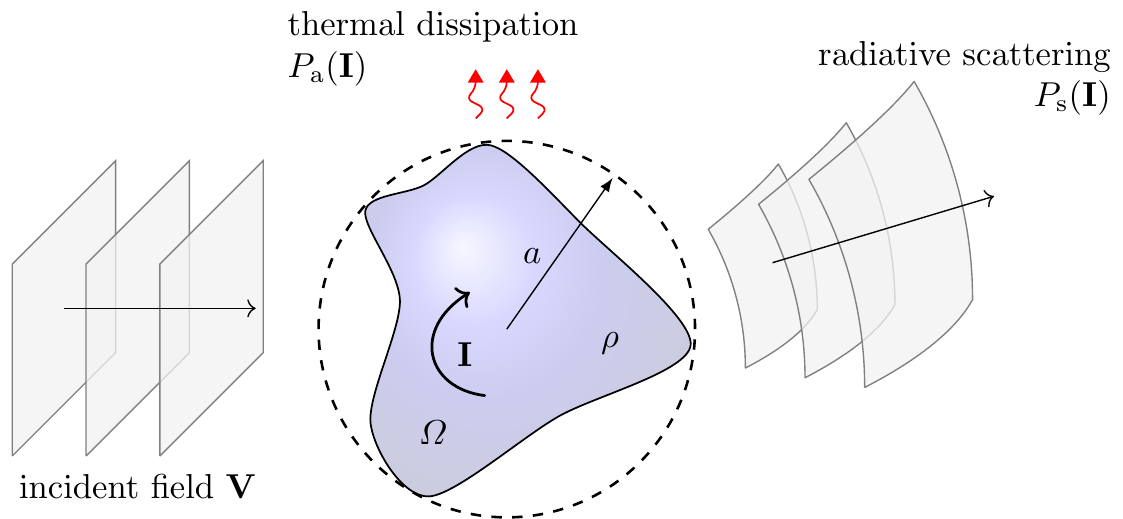}
    \caption{Schematic of arbitrary electric currents represented by the vector $\M{I}$.  The currents are confined to a region $\varOmega$ with prescribed complex resistivity $\rho$ and arbitrary shape. The region~$\varOmega$ is circumscribed by a sphere of radius~$a$.  The power extincted from an incident electric field represented by $\M{V}$ is $P_\mathrm{t}(\M{I},\M{V})$ while the scattered and absorbed powers are $P_\mathrm{s}(\M{I})$ and $P_\mathrm{a}(\M{I})$, respectively.}
    \label{fig:schematic}
\end{figure}

Assume a steady state time-harmonic electromagnetic field and consider a region $\varOmega$ containing a (possibly inhomogeneous) complex resistivity\footnote{Notice that the ``figure of merit'' used in \cite{miller2016fundamental} is equivalent to a scaled form of the real resistivity $\rho_\T{r}$.}
\begin{equation}
\rho = \rho_\mathrm{r}+\mathrm{i}\rho_\mathrm{i} 
= \frac{\eta_0\T{Im}\chi}{k|\chi|^2} + \mathrm{i} \frac{\eta_0\T{Re}\chi}{k|\chi|^2},
\label{eq:resistivity}
\end{equation}
where $\chi$ is the equivalent complex susceptibility, $\eta_0$ is impedance of free space, and $k$ is the free-space wavenumber.  When illuminated by an incident electric field $\V{E}_\mathrm{i}$, a polarization current density $\V{J}$ is induced within the region, as depicted in Fig.~\ref{fig:schematic}.  
Expanding both the induced current density and incident field into an appropriate basis, Maxwell's equations may be cast into a matrix form of the electric field integral equation
\begin{equation}
\M{V} = \M{ZI} = (\M{R}+\T{i}\M{X})\M{I},
\label{eq:vzi}
\end{equation}
where the vectors $\M{V}$ and $\M{I}$ contain expansion coefficients describing the incident field and induced current density, respectively, and $\M{Z}$ is the matrix representation of the dyadic Green's function for this particular problem, often denoted as the Method of Moments impedance matrix \cite{harrington1993field}.  Furthermore, in \eqref{eq:vzi} the matrices $\M{R}$ and $\M{X}$ are the real and imaginary components of $\M{Z}$, respectively.  By virtue of the induced current density, some of the power carried by the incident field is extincted via thermal losses and scattering, as shown in Fig.~\ref{fig:schematic}.  Note that the cycle mean absorbed power $P_\mathrm{a}$ and cycle mean scattered power $P_\mathrm{s}$ are expressible as functionals of solely the induced current $\M{I}$ \cite{gustafsson2020upper}, i.e.,
\begin{align}
\label{eq:quadforms-realA}
P_\mathrm{s}(\M{I}) &= \frac{1}{2}\M{I}^\T{H}\M{R}_0\M{I}, \\
\label{eq:quadforms-realB}
P_\mathrm{a}(\M{I}) &= \frac{1}{2}\M{I}^\T{H}\M{R}_\rho\M{I},
\end{align}
where the matrix $\M{R}_0$ is related to the Green's function operator in free space and $\M{R}_\rho$ is an operator depending on the material resistivity $\rho$.  Together, these matrices form the real part of the impedance matrix, that is,
\begin{equation}
    \M{R} = \M{R}_0+\M{R}_\rho.
\end{equation}
Functionals similar to those in \eqref{eq:quadforms-realA} and \eqref{eq:quadforms-realB} also exist for reactive energies, near-fields, and far-fields \cite{gustafsson2020upper}.

In the solution of \eqref{eq:vzi}, the induced current~$\M{I}$ is uniquely defined by the excitation field~$\M{V}$.  However, by relaxing this expression and searching over all possible polarization currents, subject to some physically-motivated constraints, bounds on the behavior of any substructure within the region $\varOmega$ may be determined \cite{gustafsson2020upper}.  For example, maximization of a quantity $C (\M{I})$ subject to real power conservation between extinction, scattering, and absorption processes may be written as an optimization problem
\begin{equation}
    \begin{aligned}
	& \myshift \max_\M{I} && C(\M{I})\\
	& \subto &&  P_\mathrm{a}(\M{I})+P_\mathrm{s}(\M{I}) = P_\mathrm{t}(\M{I},\M{V}),
\end{aligned}  
\label{eq:genOptProb-R}
\end{equation}
where
\begin{equation}
    P_\mathrm{t}(\M{I},\M{V})  = \frac{1}{2}\T{Re}\{\M{I}^\T{H}\M{V}\}
    \label{eq:quadforms-ext}
\end{equation}
is the cycle mean extincted power, i.e., the average rate of work done by the incident field upon the polarization currents. Unlike the solution to~\eqref{eq:vzi}, currents satisfying the power constraint in~\eqref{eq:genOptProb-R} are not unique.  Rather, this relaxation defines a space of currents over which an optimization procedure may search in an effort to maximize the objective functional $C(\M{I})$. This freedom also allows the current to attain zero values in subregions of~$\varOmega$.  This is equivalent to substituting those subregions with vacuum. In this way, the optimization problem~\eqref{eq:genOptProb-R}, and all optimization problems described in this paper, naturally accounts for all possible structural distributions of vacuum and the material described by the resistivity $\rho$ within the design region $\varOmega$. 

Because the functional governing absorbed power includes the material-dependent matrix $\M{R}_\rho$, the constraint in \eqref{eq:genOptProb-R} implicitly enforces behavior based on the material losses associated with the real resistivity within the design region.  For this reason, we denote this as the ``prescribed losses'' problem, with associated quantities in later sections marked with a subscript $_\M{R}$.  Adding an appropriate second constraint further restricts the optimization problem to enforce conservation of reactive power, based on the imaginary component of the design region resistivity,
\begin{equation}
    \begin{aligned}
	& \myshift \max_\M{I} && C(\M{I})\\
	& \subto &&  P_\mathrm{a}(\M{I})+P_\mathrm{s}(\M{I}) = P_\mathrm{t}(\M{I},\M{V})\\
	& &&  W(\M{I}) = W_\mathrm{t}(\M{I},\M{V})
\end{aligned}  
\label{eq:genOptProb-Z}
\end{equation}
where $W$ and $W_\mathrm{t}$ are functionals associated with reactive (complex) power flow, see~\cite{gustafsson2020upper} for their explicit forms resembling~\eqref{eq:quadforms-realA}, \eqref{eq:quadforms-realB} and~\eqref{eq:quadforms-ext}.  Because the two constraints together enforce both real and imaginary components of the design region's material properties, we refer to this as the ``prescribed materials'' case and use a subscript $_\M{Z}$ to distinguish these results from the prescribed losses case. 

Though the problems in \eqref{eq:genOptProb-R} and \eqref{eq:genOptProb-Z} are applicable to arbitrary excitation fields and a wide variety of objective functionals, we restrict our focus here toward the study of functionals consisting of combinations of absorbed and scattered powers under plane wave incidence.  As such, throughout this paper we normalize all optimized power quantities to their respective cross section $\sigma = P/S_0$, where $S_0$ denotes the incident plane wave power density.

\begin{figure}
    \centering
    \includegraphics[width=3.3in]{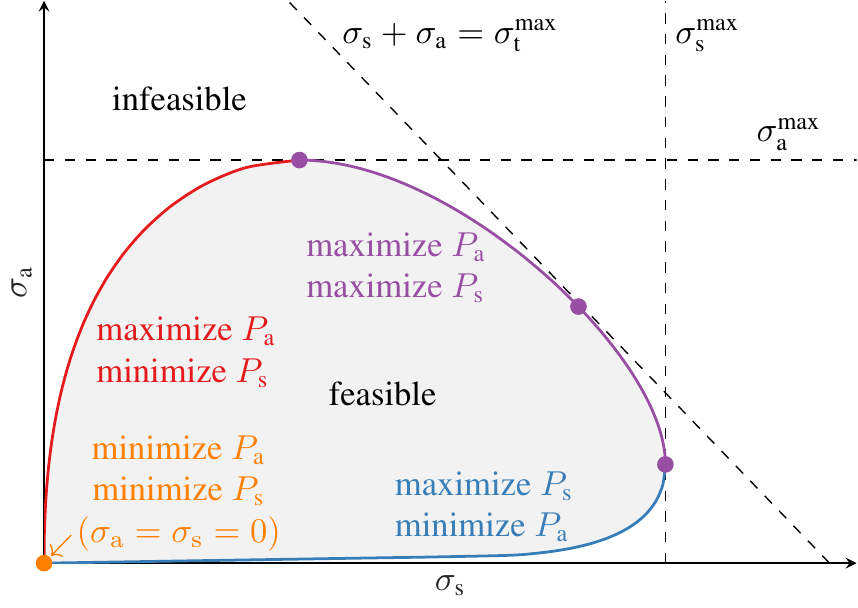}
    \caption{Schematic of a Pareto front between absorption and scattering.
The Pareto optimal set is formed by three parts, the description of which
is shown along the curves. Feasible solutions may exist only on the
front or within the shaded region. The maximal extinction cross section
$\sigma_\mathrm{t}^\mathrm{max}$ always lies on the Pareto front at the
point with maximum $L_1$ ($||\cdot||_1$) norm given by
$\sigma_\mathrm{a}+\sigma_\mathrm{s}$.}
    \label{fig:pareto}
\end{figure}

Under either one or two constraints, bounds on absorption, scattering, and extinction cross sections take the form of \eqref{eq:genOptProb-R} or \eqref{eq:genOptProb-Z}.  Each of these problems optimizes a single parameter in isolation, yielding an absolute upper bound, see dashed lines in in Fig.~\ref{fig:pareto}.  It is, however, unknown whether scattering and absorption can be simultaneously optimized, i.e., whether the intersection of the dashed lines may be reached.  Additionally,  rather than seeking simultaneous maximization of both absorption and scattering, we may pose similar questions regarding the feasibility of structures maximizing absorption while minimizing scattering (and vice versa).

In subsequent sections, we formulate multiobjective optimization problems \cite{1978CohonMultiobjectiveProgrammingAndPlanning,Deb_MultiOOusingEA} to study these problems in detail. The results of these analyses lead to a feasible objective space, shown as a shaded region in Fig.~\ref{fig:pareto}, where all physically realizable responses must lie.  The boundary of this region, drawn as the thick curve in Fig.~\ref{fig:pareto}, is defined as the set of points at which an increase in performance (defined either as maximization or minimization) in one parameter must be accompanied by a performance decrease in the other. 

The boundary of the feasible region may be constructed by several means.  Here we conceptually separate this boundary into three Pareto optimal sets. The first (purple thick line in Fig.~\ref{fig:pareto}) is the simultaneous maximization of both absorption and scattering. Note that maximal extinction, defined as the sum of absorption and scattering, is always located on this portion of the feasible region boundary. The second boundary segment (red thick line in Fig.~\ref{fig:pareto}) embodies minimization of scattering and maximization of absorption. Finally, there is a Pareto optimal set for minimum absorption and maximum scattering (blue thick line in Fig.~\ref{fig:pareto}). Minimization of both parameters (orange marker in Fig.~\ref{fig:pareto}) collapses to a point at the origin and does not represent a meaningful Pareto optimal set for this particular class of problems.

\section{Prescribed losses}
\label{sec:losses}

Under the constraint of real power conservation, the boundary of the feasible objective space in Fig.~\ref{fig:pareto} is formed by the union of Pareto fronts arising from two multiobjective optimization problems.  The first maximizes absorption while maintaining a fixed scattering cross section~\cite{2008EichfelderAdaptiveScalarizationMethodsInMultiobjectiveOptimization}, i.e.,
\begin{equation}
\begin{aligned}
& \myshift \max && P_\mathrm{a} \\
&\mathrm{s.t.} && P_\mathrm{a}+P_\mathrm{s} - P_\mathrm{t} = 0 \\
& && P_\mathrm{s} \in \left[\min\{ P_\mathrm{s} \}, \max\{ P_\mathrm{s} \} \right].
\end{aligned}
\label{eq:const-r-pa}
\end{equation}
Solution of this problem yields the red and purple curves in Fig.~\ref{fig:pareto}.  The second problem reverses the roles of absorption and scattering,
\begin{equation}
\begin{aligned}
& \myshift \max && P_\mathrm{s} \\
&\mathrm{s.t.} && P_\mathrm{a}+P_\mathrm{s} - P_\mathrm{t} = 0 \\
& && P_\mathrm{a} \in \left[\min\{ P_\mathrm{a} \}, \max\{ P_\mathrm{a} \} \right],
\end{aligned}
\label{eq:const-r-ps}
\end{equation}
yielding the purple and blue curve in Fig.~\ref{fig:pareto}.  Either of the problems in \eqref{eq:const-r-pa} or \eqref{eq:const-r-ps} may be rewritten as minimizations by changing the sign of the objective functional.

While these problems are straightforward to implement, they can be almost entirely replaced by a weighted problem~\cite{1978CohonMultiobjectiveProgrammingAndPlanning,Deb_MultiOOusingEA} under a single constraint, i.e.,
\begin{equation}
\begin{aligned}
& \myshift \max && w_\T{a} P_\mathrm{a}+w_\mathrm{s}P_\mathrm{s} \\
&\mathrm{s.t.} && P_\mathrm{a}+P_\mathrm{s} - P_\mathrm{t} = 0,
\label{eq:loss-weights}
\end{aligned}
\end{equation}
where $w_\T{a}$ and $w_\T{s}$ are arbitrary real weights.  One slight disadvantage of the method of weights employed in \eqref{eq:loss-weights} is that it assumes a concave / convex Pareto front \cite{2008EichfelderAdaptiveScalarizationMethodsInMultiobjectiveOptimization}.  Hence this method fails to produce any linear portions of the feasible region boundary, occurring in one special case discussed later in this section.  Note that the portion of the boundary maximizing both absorption and scattering (purple curve in Fig.~\ref{fig:pareto}) is achieved by the weights $w_\T{a} = \alpha$, $w_\T{s} = (1-\alpha)$ for $\alpha \in [0,1]$.  Additionally, the single objective optimization problems in absorption, scattering, and extinction are achieved by specific values of weights, specifically $w_\T{s}=0$, $w_\T{a}=0$, and $w_\T{s}=w_\T{a}$.  For clarity, the remainder of this section keeps both weights $w_\T{a}$ and $w_\T{s}$ as independent parameters, though the full range of their ratios can easily be covered by their parameterization in a single parameter, e.g., $w_\T{a} = \cos\phi$, $w_\T{s} = \sin\phi$, with $\phi\in[0,2\pi]$.

Explicitly inserting the quadratic and linear forms for each of the above quantities in \eqref{eq:loss-weights}, we obtain the quadratically constrained quadratic program (QCQP),
\begin{equation}  
\begin{aligned}
& \myshift \max_\Im && \Im^\mathrm{H}(w_\T{a} \Rm_{\rho}+w_\T{s}\Rm_{0}){\Im} \\
&\mathrm{s.t.} &&  \Im^\mathrm{H}(\Rm_{\rho}+\Rm_{0})\Im-\mathrm{Re}\{\Im^\mathrm{H}\Vm\} = 0.
\label{eq:primal-losses}
\end{aligned}
\end{equation}
For any set of weights, the optimization problem is solved by minimizing the associated convex dual function \cite[App.~B]{boyd2004convex},\cite[App.~B]{gustafsson2020upper}, i.e.,
\begin{equation}
    \min_{\nu\in\mathcal{D}}g(\nu) = \min_{\nu\in\mathcal{D}}  \nu^{2}\Vm^\mathrm{H}((\nu-w_\T{a})\Rm_{\rho}+(\nu-w_\T{s})\Rm_{0})^{-1} \Vm  .
    \label{eq:dual-r}
\end{equation}
The domain $\mathcal{D}$ restricts the dual parameter $\nu$ to values ensuring positive definiteness of the matrix inverted in~\eqref{eq:dual-r},
\begin{equation}
    \mathcal{D} = \{\nu : (\nu-w_\T{a})\Rm_{\rho}+(\nu-w_\T{s})\Rm_{0} \succ 0 \}.
\end{equation}

Solution of the dual minimization problem is greatly simplified by considering the radiation mode eigenvalue decomposition~\cite{schab2016modal}\cite[App.~C]{gustafsson2020upper}
\begin{equation}
    \Rm_{0}\Im_{n} = \varrho_{n}\Rm_{\rho}\Im_{n}
    \label{eq:rad-mode-gep}
\end{equation}
and the basis it generates.  Applying this expansion to diagonalize all matrices in \eqref{eq:dual-r}, we find the range of the dual parameter
\begin{equation}
    \mathcal{D} = (\nu_\T{min},\infty],
\end{equation}
where
\begin{equation}
    \nu_\T{min} = \max_{n}\frac{w_\T{s}\varrho_{n}+w_\T{a}}{1+\varrho_{n}},
\end{equation}
and reduce the quadratic form~\eqref{eq:dual-r} to a simple summation
\begin{equation}
    \min_{\nu\in\mathcal{D}}g(\nu) = \min_{\nu>\nu_\T{min}}\nu^{2}\sum_{n=1}^{N} \frac{|\hat{V}_{n}|^{2}}{(\nu-w_\T{a})+(\nu-w_\T{s})\varrho_{n}},
    \label{eq:sum-r}
\end{equation}
where $\tilde{\M{V}} = \M{Q}^\T{H}\M{V}$ and $\M{Q}$ is the matrix with eigenvectors satisfying \eqref{eq:rad-mode-gep} as its columns.  Solving~\eqref{eq:sum-r} gives the optimal dual parameter value $\nu^*$, from which the optimal current $\Im^*$ may be calculated as
\begin{equation}
    \Im^*= \frac{\nu^*}{2}((\nu^*-w_\T{a})\Rm_{\rho}+(\nu^*-w_\T{s})\Rm_{0})^{-1} \Vm.
    \label{eq:r-iopt}
\end{equation}
For each set of weights $w_\T{a}$ and $w_\T{s}$ , the optimal absorbed and  scattered powers $P_\mathrm{a}(\Im^*)$ and $P_\mathrm{s}(\Im^*)$ may be calculated, normalized to cross sections, and plotted to form the Pareto front shown in Fig.~\ref{fig:pareto}.

\subsection{Minimal absorption with maximal scattering}
\label{sec:losses-minmax}
For isotropic resistivities with positive, non-zero loss throughout the region of interest, the operator~$\M{R}_\rho$ is not compact, leading to a full rank matrix representation.  However, the scattering operator $\M{R}_0$ is compact, leading to an accumulation of eigenvalues at zero in the radiation eigenmode problem \eqref{eq:rad-mode-gep}.  We can interpret the difference in operator structures as an indication that there exists currents which absorb arbitrarily well and scatter arbitrarily poorly, though the converse case is not necessarily true.  Practically, this asymmetry leads to a slight technical difficulty in computing the portion of the feasible objective space boundary in the case where scattering is maximized while absorption is heavily penalized (see portion of blue curve in Fig.~\ref{fig:pareto} closest to the origin). 

To examine this issue in detail, we begin by noting that, because of the leading $\nu^2$ term, the dual function $g(\nu)$ in \eqref{eq:dual-r} is minimized at a value $\nu^*>0$ when $\nu_\T{min}>0$ and is always minimized at $\nu^*=0$ when $\nu_\T{min}<0$.  Solutions where $\nu^*=0$ yield identically zero currents with zero scattering and absorption cross sections.  We now examine the set of weights maximizing scattering while penalizing absorption where $\nu_\T{min}\rightarrow 0^+$ when $w_\T{a}<0$ and $w_\T{s}>0$.  For simplicity, consider the parameter $w_\T{s}$ to be fixed. Under this set of conditions, the lower limit of the dual parameter domain $\nu_\T{min}$ approaches zero when
\begin{equation}
    w_\T{a} \rightarrow (-w_\T{s}\varrho_1)^+,
\end{equation}
with $\varrho_1$ being the largest radiation mode eigenvalue.  By \eqref{eq:dual-r}, the optimal dual parameter $\nu^*$ approaches zero in this limit (i.e., as $\nu_\T{min}$ approaches zero from above).  Substituting these features into the current associated with the dual function using the optimal dual parameter $\nu^*$ yields
\begin{equation}
    \lim_{w_\T{a}\rightarrow (-w_\T{s}\varrho_1)^+}\M{I}^*\approx -\frac{w_\T{s}\nu^{*}}{2}\left(\M{R}_0 - \varrho_1^-\M{R}_\rho\right)^{-1}\M{V},
\end{equation}
where $\varrho_1^-$ denotes a value approaching $\varrho_1$ from below.  The radiation eigenmode problem simplifies the matrix inverse, leading to
\begin{equation}
    \lim_{w_\T{a}\rightarrow (-w_\T{s}\varrho_1)^+}\M{I}^*\approx -\frac{w_\T{s}\nu^{*}}{2}\sum_n\Im_{n}\frac{\tilde{V}_n}{\varrho_n-\varrho_1^-}\sim \M{I}_1
    \label{eq:i-approx}
\end{equation}
where $\M{I}_1$ is the eigenvector associated with $\varrho_1$ and it is assumed that the incident field has a non-zero projection onto this mode, i.e., $\tilde{V}_1 \neq 0$.  By the very nature of the radiation mode eigenvalue problem, the coordinate in $\sigma_\T{s} / \sigma_\T{a}$ space corresponding to this current has the feature $\sigma_\T{s} = \rho_1\sigma_\T{a}$.  The solution current in \eqref{eq:i-approx} may be altered by any phase satisfying 
\begin{equation}
    0 \leq \T{Re}\{\mathrm{e}^{\mathrm{j}\theta}\M{I}_1^{\T{H}}\M{V}\} \leq \T{Re}\{\M{I}_1^\T{H}\M{V}\}
\end{equation}
while maintaining the condition $\sigma_\T{s} = \rho_1\sigma_\T{a}$.  In this way the line segment forming a straight part of the feasible region boundary between the point where $\nu_\T{min}\rightarrow0^+$ and the origin is formed.

\subsection{Electrically small limits}
\label{sec:losses-esa}

Several simplifications can be made in the long wavelength limit.  We begin by studying the low frequency behavior of the simultaneous maximization component of the feasible objective boundary by setting the weights $w_\T{a}=\alpha$, $w_\T{s} = 1-\alpha$ with $\alpha \in [0,1]$. In the limit of vanishingly small electrical size $ka \ll 1$,~$a$ being the radius of the smallest sphere circumscribing the scatterer, the sum in \eqref{eq:sum-r} is dominated by the first term\footnote{Numerically, the first two or three modes may need to be included depending on dimensionality and ``near'' degeneracies, but this will not change the form of the following results.} (largest $\varrho_n$) \cite{gustafsson2020upper}.  Keeping only the dominant $n = 1$ term, whose eigenvalue~$\varrho_1$ may be approximated based solely on frequency and the scatterer's volume $V$ and real resistivity
\begin{equation}
    \varrho_1 \approx \frac{\eta_0 k^2 V}{6\pi \rho_\mathrm{r}},
\end{equation}
the minimization in \eqref{eq:sum-r} can be solved analytically, yielding the optimal dual parameter
\begin{equation}
    \nu_\alpha^* \approxkazero \frac{2(\alpha(1-\varrho_1)+\varrho_1)}{1+\varrho_1}.
\end{equation}
From \eqref{eq:r-iopt}, the optimal current is then found to be independent of the parameter~$\alpha$, i.e.,
\begin{equation}
    \M{I}_\alpha^* \approxkazero \M{I}_1\frac{\tilde{V}_1}{1+\varrho_1}.
\end{equation}
Hence, in the electrically small limit, there is no trade-off between maximal absorption and scattering and both parameters are maximized by the same current density, namely the dominant radiation mode. Inserting this current into the quadratic forms for scattering and absorption cross-sections we see the familiar low-frequency forms~\cite{gustafsson2020upper}
\begin{subequations}
\begin{equation}\label{esl-abs}
    \sigma_{\mathrm{a},\alpha} \approxkazero \frac{6\pi}{k^2}\frac{\varrho_1}{(1+\varrho_1)^2},
\end{equation}
\begin{equation}\label{esl-scat}
        \sigma_{\mathrm{s},\alpha} \approxkazero \frac{6\pi}{k^2}\frac{\varrho_1^2}{(1+\varrho_1)^2}.
\end{equation}
\label{eq:r-esa-cacs}
\end{subequations}
Graphically, this result indicates that in the electrically small limit, the portion of the feasible objective boundary corresponding to simultaneous maximization of absorption and scattering (purple curve, Fig.~\ref{fig:pareto}) collapses to a single point.  Because the dominant radiation mode represents the highest possible ratio of scattering relative to absorption, the portion of the boundary representing minimum absorption and maximum scattering (blue curve, Fig.~\ref{fig:pareto}) in this case becomes a straight line between the origin and the point described by \eqref{eq:r-esa-cacs}.  The same extreme simplification does not apply for the case of maximizing absorption while minimizing scattering (red curve, Fig.~\ref{fig:pareto}), though by observing that in the electrically small limit $\rho_n \ll \rho_1$ for $n>2$, the minimization problem in \eqref{eq:sum-r} may be reduced to
\begin{equation}
    \min_{\nu\in\mathcal{D}}g(\nu) = \min_{\nu>w_\T{a}}\nu^{2}\left[ \frac{\sum_{n=2}^{N}|\hat{V}_{n}|^{2}}{(\nu-w_\T{a})}+\frac{|\hat{V}_{1}|^{2}}{(\nu-w_\T{a})+(\nu-w_\T{s})\rho_1}\right],
    \label{eq:sum-r-esa-mincs-maxca}
\end{equation}
the solution of which is obtained by finding the roots of a cubic polynomial.

\subsection{Example: maximal trade-offs with prescribed losses}

As an example calculation, several Pareto fronts for maximizing scattering and absorption cross sections are calculated for a spherical region of radius $a$ over a range of electrical sizes~$ka$ using several real resistivities~$\rho_\mathrm{r}$.  In this example and throughout the remainder of this paper, we normalize real resistivities by the dimension $a$ of the design region, i.e., $\rho/a$.  This could be further scaled by the free space impedance $\eta_0$ and the unitless electrical size of the design region $ka$, to give the figure of merit \cite{miller2016fundamental}
\begin{equation}
    \frac{\rho_\T{r}}{a}\frac{ka}{\eta_0} = \frac{k\rho_\T{r}}{\eta_0} = \frac{\T{Im}~\chi}{|\chi|^2}.
\end{equation}  

In addition to the Pareto fronts (black curves), the points of maximum absorption, extinction, and scattering are marked with solid green, blue, and red lines, respectively.  Decade values of electrical size $ka$ are marked with thicker curves to illustrate the trend of increasing electrical size. Moreover, the electrically small limit is plotted as a dashed line to graphically show its approximation relative to the Pareto-optimal set. 

\begin{figure}
    \centering
    \includegraphics[width=5in]{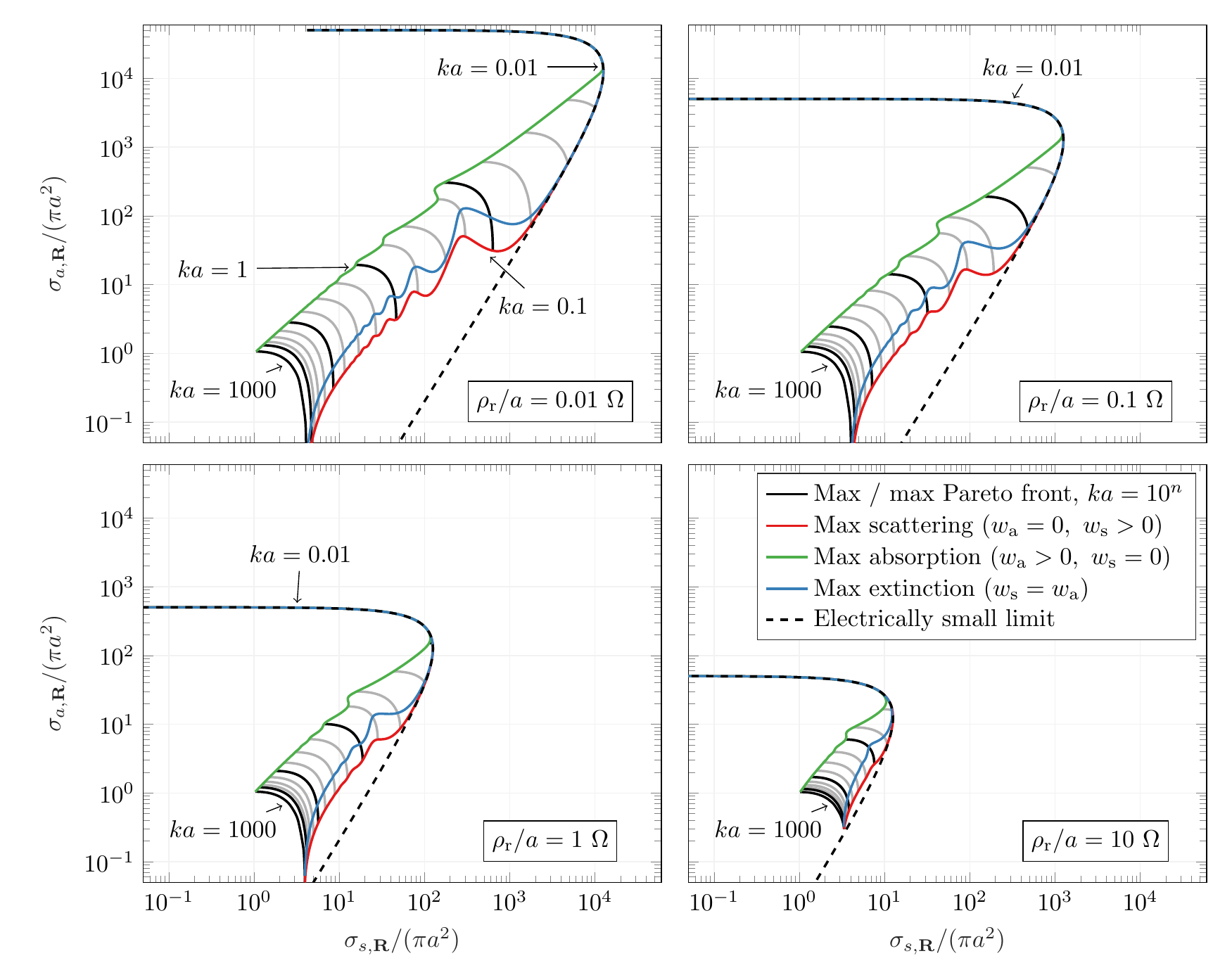}
    \caption{Trade-offs between optimal absorption and scattering under a real power conservation constraint. The support~$\varOmega$ of the optimized current is a sphere of electrical size $ka\in[0.01,~1000]$ filled with material of different real resistivities~$\rho_\mathrm{r}/a$.}
    \label{fig:PL_four-panel}
\end{figure}

As electrical size decreases, the set of optimal solutions to \eqref{eq:primal-losses} converge to a single point, seen graphically by the collapse of the Pareto fronts in Fig.~\ref{fig:PL_four-panel}. This implies that for sufficiently small electrical size $ka$, the Pareto optimal set contains values which simultaneously represent maximization of absorption, scattering, and extinction, in line with derivations provided in the previous section. 
Furthermore, the traces corresponding to maximum extinction (blue) and maximum scattering (red) converge to the electrically small limit more rapidly than that of maximal absorption. At small electric sizes, this phenomenon occurs as maximum extinction is dominated by maximum scattering.  For large electrical sizes, the normalized maximum absorption and scattering cross sections converge to 1 and 4, respectively, for all values of real resistivity.

As resistivity increases, the Pareto-optimal set converges to a single point for increasingly larger electrical size $ka$. For sufficiently large electrical size and small real resistivity $\rho_\mathrm{r}$, the trade-off curve becomes more pronounced with a bias towards scattered power. Additionally, for small electrical size, the electrically small limit closely follows the maximum scattering curve.

\subsection{Example: feasible objective spaces as functions of size and loss}
\begin{figure}
    \centering
    \includegraphics[width=2.25in]{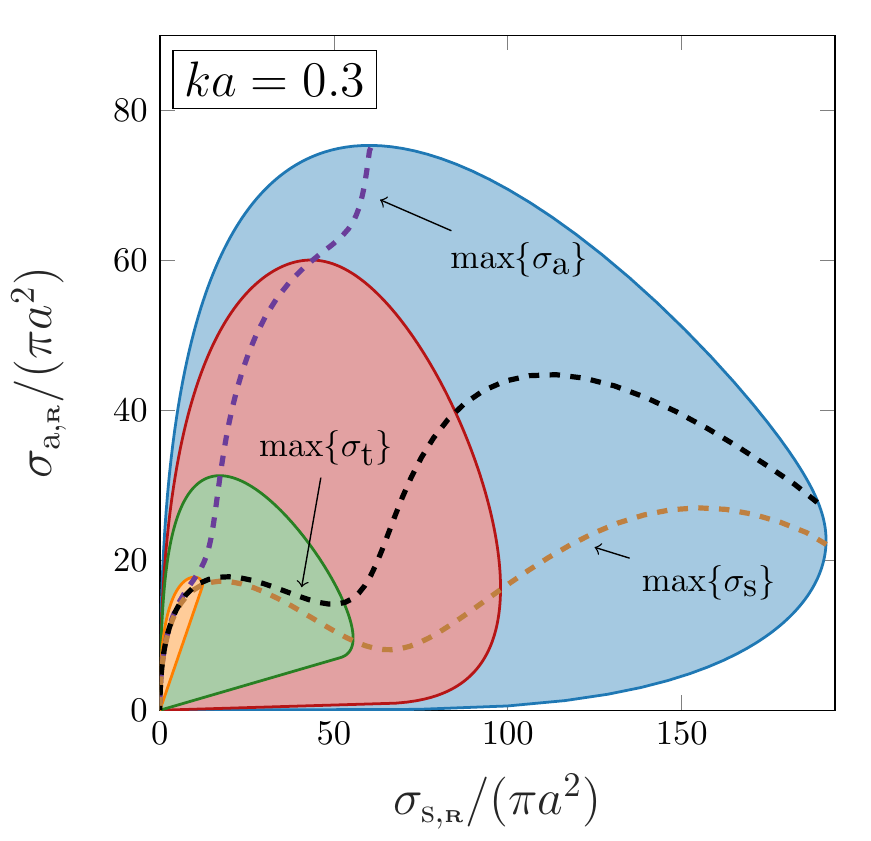}
    \includegraphics[width=2.25in]{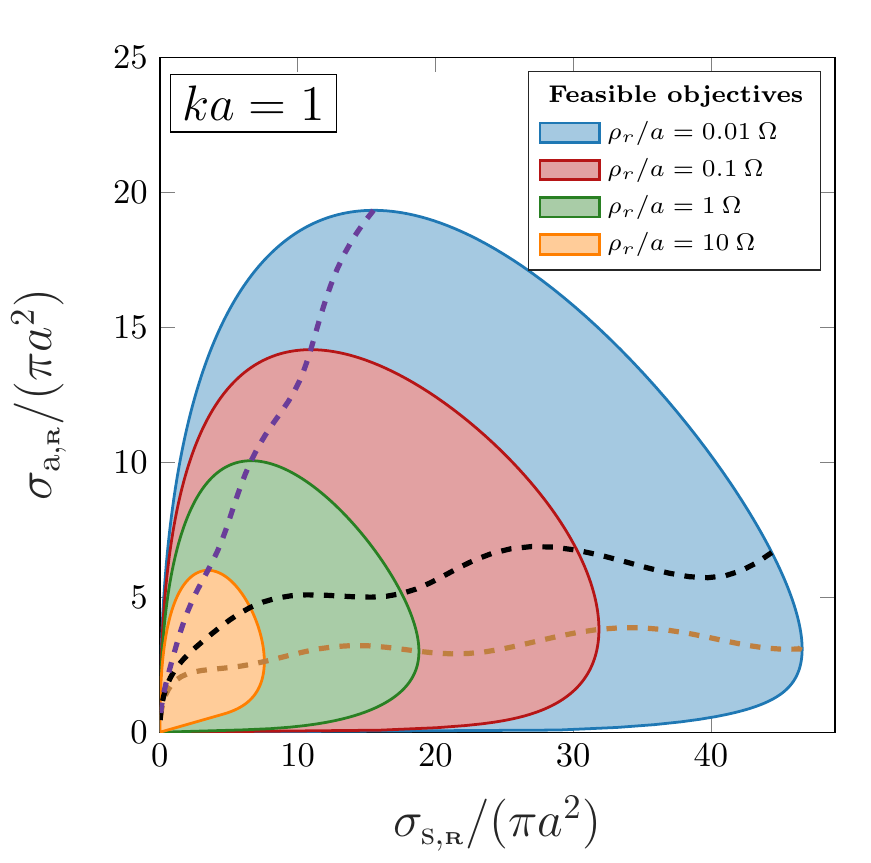}
    \caption{Feasible sets on $\sigma_{\mathrm{s},\tiny{\textbf{R}}}/(\pi a^2)$, $\sigma_{\mathrm{a},\tiny{\textbf{R}}}/(\pi a^2)$ for obstacles composed of a material with $\rho_\mathrm{r}/a\in\{0.01,0.1,1,10\}\unit{\Omega}$ circumscribed by spheres with radii $ka=0.3$ (left), and $ka = 1$ (right). }
    \label{fig:r-feasible}
\end{figure}

To examine the entire feasible objective space under the prescribed losses constraint, several closed Pareto fronts are generated for two electrical sizes $ka$ and various fixed real resistivites $\rho_r$, see Fig.~\ref{fig:r-feasible}.  In a similar fashion to the case investigating maximal trade-offs in Fig.~\ref{fig:PL_four-panel}, the portions of Pareto fronts corresponding to maximal scattering and absorption cross-sections converge to a single point as resistivity increases. The feasible objective space is limited by the upper bounds on maximal absorption and scattering cross-sections, with the space growing as the maximal cross-sections take on larger values. 

\subsection{Example: performance of simple geometries in comparison to feasible objective spaces}
\label{sec:losses-realized}

\begin{figure}
    \centering
    \includegraphics[height=2.75in]{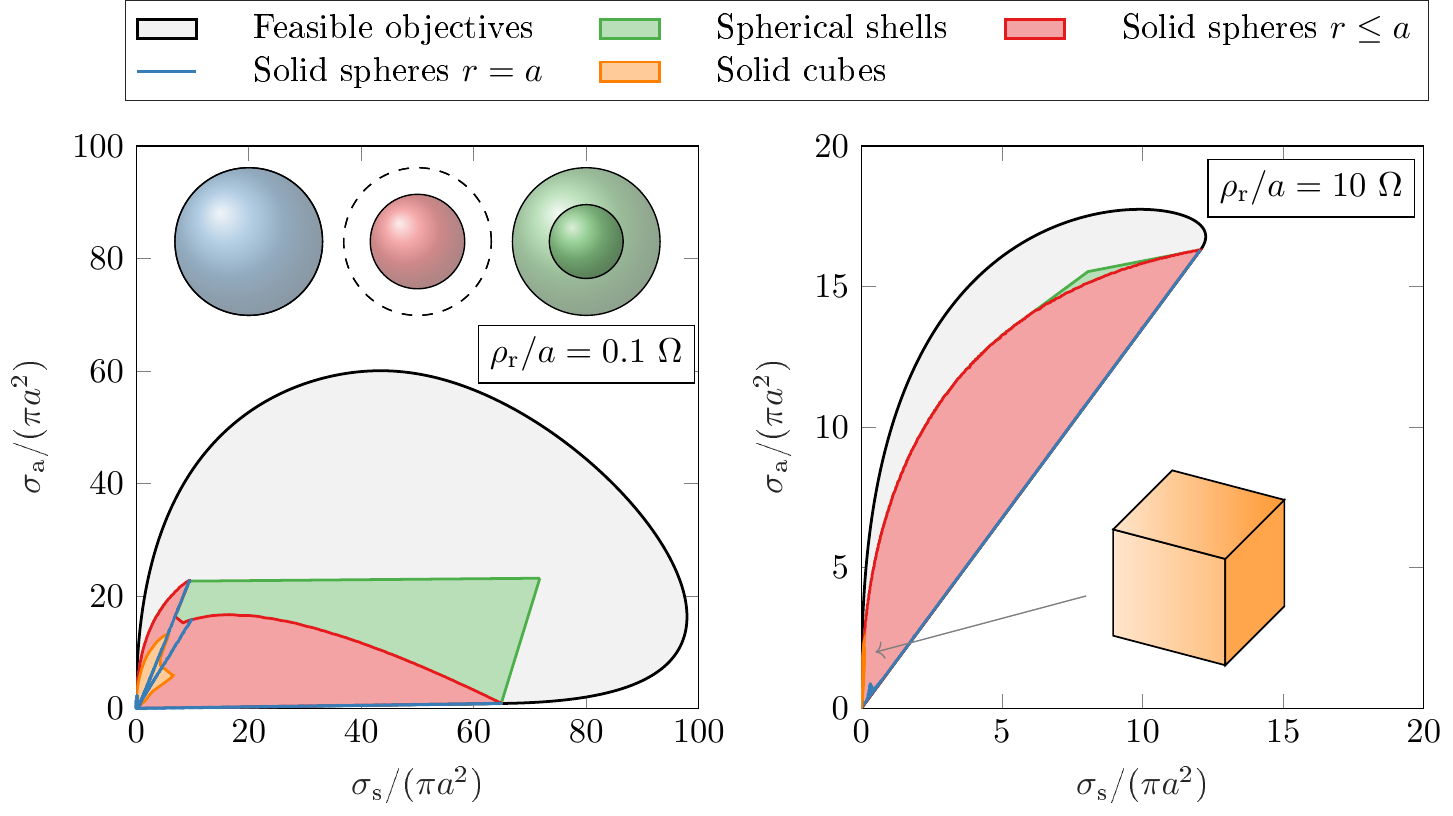}  
    \caption{Realized cross-sections for designs confined to a sphere of size $ka=0.3$ with real resistivity $\rho_\mathrm{r}/a=0.1~\Omega$ (left) and $\rho_\mathrm{r}/a=10~\Omega$ (right).}
    \label{fig:r-realized}
\end{figure}

In Fig.~\ref{fig:r-realized}, we examine the ability of simple objects to span the feasible solution region produced by a spherical design space of electrical size $ka = 0.3$ in a low-loss, $\rho_\mathrm{r}/a = 0.1~\Omega$, and high-loss, $\rho_\mathrm{r}/a = 10~\Omega$, configurations.  In each case, the feasible solution region is denoted by a gray color, while the realized absorption and scattering cross-sections of various parameterized structures are denoted by blue, red, green, and orange colors.

Solid spheres of radius $r=a$ are parameterized by their imaginary resistivity $\rho_\T{i}$ and trace out the blue locus in Fig.~\ref{fig:r-realized}.  They achieve nearly optimal performance in terms of maximizing scattering while minimizing absorption, though most of the feasible objective space is unreachable by this particular structure.  Solid spherical objects parameterized by both imaginary resistivity and radius $r\leq a$ (red color) span more significant portion of the feasible region in both loss scenarios, particularly in the high-loss case, where they achieve nearly optimal performance. This is not the case in the low resistivity case, where solid spheres span only a small portion of the feasible region. Core-shell structures with fixed outer radius~$r = a$, core radius $r' = a/2$ and independently varying imaginary resistivity in both core and shell layers extend further into the feasible objective region, though this extension is much more significant in the low-loss case.

The feasible objective space in Fig.~\ref{fig:r-realized} bounds the performance of arbitrarily complex structures within the prescribed design region, not just those with spherical symmetry.  As a demonstration of this feature, the realized scattering and absorption cross-sections of solid cube structures with varying size and imaginary resistivity are calculated and plotted (orange color) in both panels of Fig.~\ref{fig:r-realized}. Solutions for this realized structure were obtained via volumetric method of moments \cite{polimeridis2014stable}. In both cases, the incident field is polarized along the normal of one pair of the cube's faces.  The realized cross-sections are well within the bounding feasible objective space, providing significantly smaller responses than optimized spherical structures.

\section{Prescribed materials}
\label{sec:materials}

The multi-objective optimization of scattering and absorption cross sections can be extended to the prescribed materials case by adding a second constraint related to the reactive energy of the system.  Substituting in the explicit forms of all objectives and constraints, the following optimization problem is obtained
\begin{equation}
\begin{aligned}
&\myshift\max_\Im && \Im^\mathrm{H}(w_\T{a} \Rm_{\rho}+w_\T{s}\Rm_{0}){\Im} \\
&\mathrm{s.t.} &&  \Im^\mathrm{H}(\Rm_{\rho}+\Rm_{0})\Im-\mathrm{Re}\{\Im^\mathrm{H}\Vm\} = 0\\
& &&  \Im^\mathrm{H}\Xm\Im-\mathrm{Im}\{\Im^\mathrm{H}\Vm\} = 0.
\end{aligned}
\end{equation}
Using similar techniques as in previous section, the problem is solved by minimizing the dual function
\begin{equation}
    \min_{\nu,\mu\in\mathcal{D}} g(\nu,\mu) = \\ \min_{\nu,\mu\in\mathcal{D}}  \frac{\nu^{2}+\mu^{2}}{8S_0}\Vm^\mathrm{H}((\nu-w_\T{a})\Rm_{\rho}+(\nu-w_\T{s})\Rm_{0}+\mu\Xm)^{-1} \Vm  ,
    \label{eq:dual-prescribed}
\end{equation}
where the domain of Lagrange parameters $\nu$ and $\mu$ is restricted to
\begin{equation}
    \mathcal{D} = \{(\nu,\mu) : (\nu-w_\T{a})\Rm_{\rho}+(\nu-w_\T{s})\Rm_{0}+\mu\Xm \succ 0 \} .
\end{equation}
Because \eqref{eq:dual-prescribed} involves a parameterized matrix inverse involving three matrices, it cannot be reduced to a summation similar to \eqref{eq:sum-r}.  An exception occurs when $w_\T{a}=w_\T{s}$, corresponding to maximization of extinction cross section \cite{gustafsson2020upper}.

\subsection{Example: dielectric scatterers}
\label{mat-diel}

\begin{figure}
    \centering
    \includegraphics[width=3.5in]{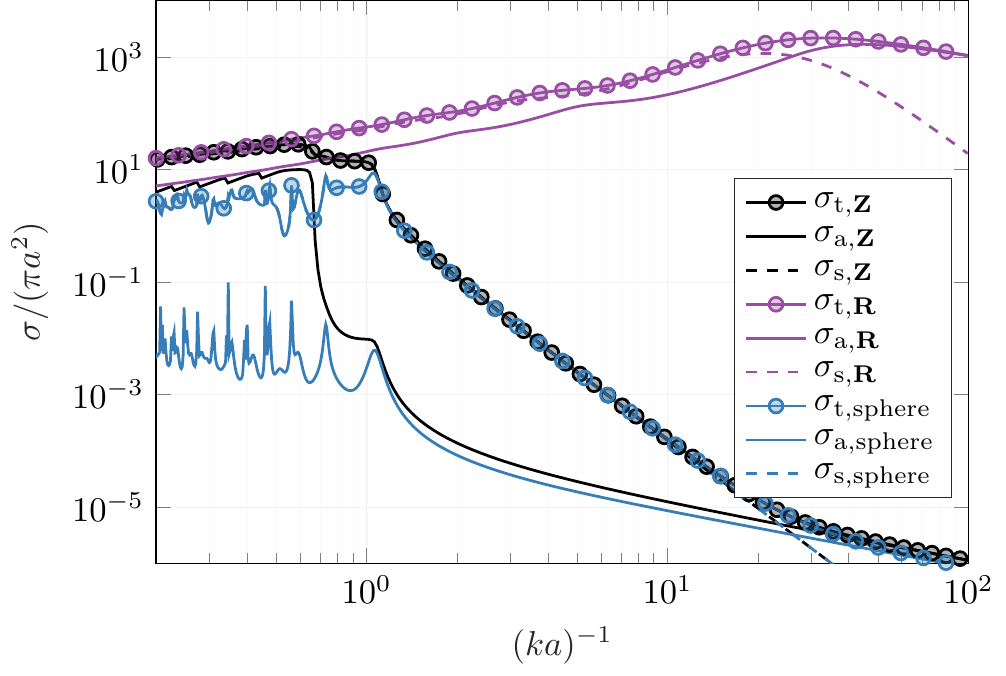}
    \caption{Example calculation highlighting impact of reactance constraint. Purple curves correspond to prescribed losses constraint, black curves correspond to prescribed materials constraints, and blue curves correspond to realized cross-sections of a homogeneous sphere filled with permittivity $\epsilon_\mathrm{r} = 10 + \mathrm{i}10^{-3}$.}
    \label{fig:dielectric-individual}
\end{figure}

As a first example, we consider a lossy dielectric sphere with relative permittivity \mbox{$\varepsilon_\mathrm{r} = 10+\mathrm{i}10^{-3}$}.  In Fig.~\ref{fig:dielectric-individual}, single objective bounds on scattering ($w_\T{a} = 0$, $w_\T{s} > 0$), absorption ($w_\T{a} > 0$, $w_\T{s} = 0$), and extinction ($w_\T{a} = w_\T{s}$) are plotted as functions of electrical size~$ka$ under both prescribed losses and prescribed materials constraints.  The realized cross-sections of a homogeneous sphere driven by the incident field are also plotted.  We observe that the addition of the prescribed materials constraint severely restricts all maximal cross sections at low frequencies (small electrical size), though at high frequencies the two sets of bounds take on similar values.  

The realized cross sections of a homogeneous sphere are very close to the prescribed materials bounds for electrical sizes $ka<1$, indicating that this very simple geometry is nearly ideal in the long wavelength limit.  It is thus possible to conclude that for scattering or absorption enhancement in the low-frequency regime, advanced inverse design routines would provide little advantage over the simple homogeneous sphere.  This result, however, is specific to this particular material, as seen in later examples involving metallic objects.

\begin{figure}
    \centering
    \includegraphics[height=2.25in]{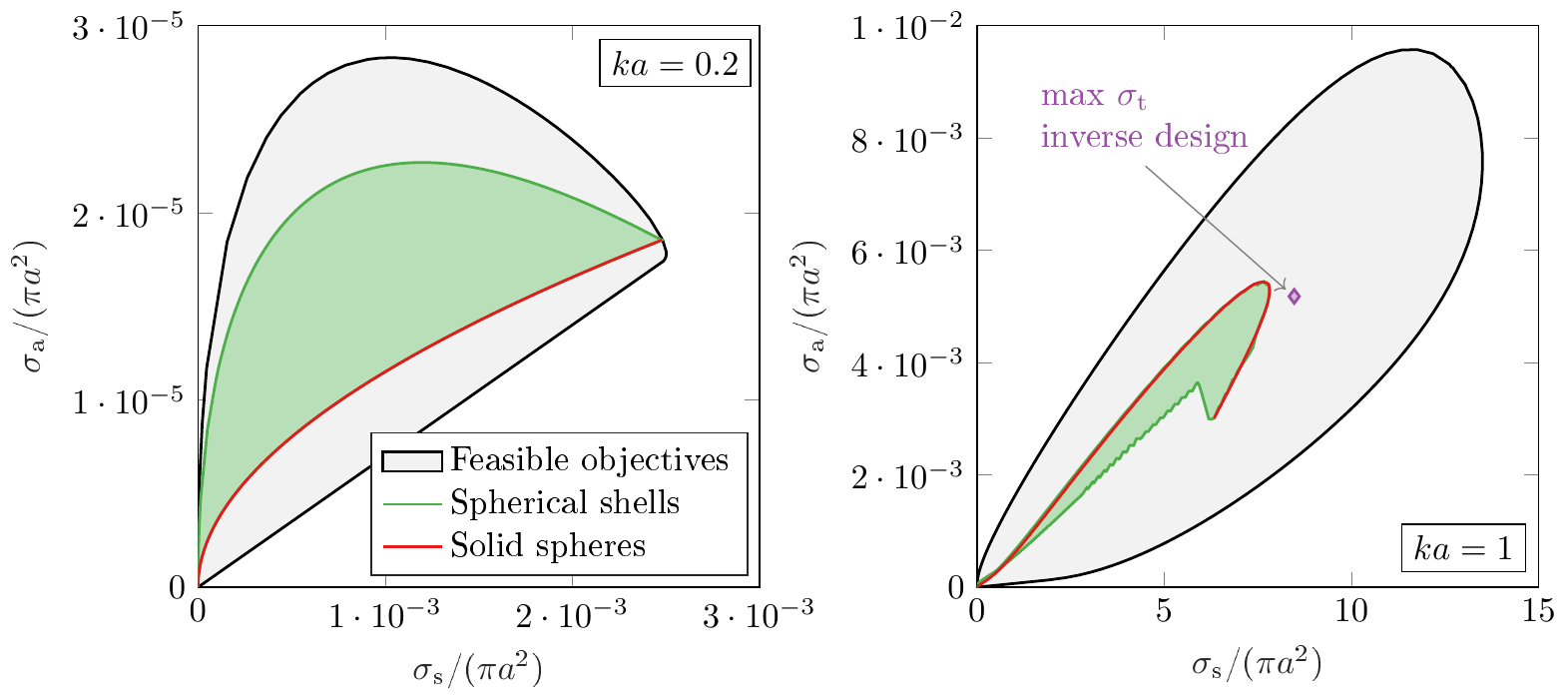}
    \caption{Realized cross-sections for designs confined to a sphere with electrical size $ka = 0.2$ (left) and $ka = 1$ (right) constructed of dielectric material with relative permittivity $\varepsilon_\T{r} = 10+\T{i}10^{-3}$, overlaid on the feasible region calculated with prescribed materials constraints. Realized designs are solid spheres of radius~$r \leq a$ (red locus) and spherical shells with radii~$r_1,r_2 \leq a$ (green shaded region).}
    \label{fig:dielectric-feasible}
\end{figure}

Feasible regions for this configuration, along with the realized cross sections of homogeneous spheres with swept radius (red locus) and spherical shells with swept inner and outer radii (green shaded region) are shown in Fig.~\ref{fig:dielectric-feasible}.  For the smaller electrical size ($ka = 0.2$, left panel), gains in absorption can be achieved by the use of a spherical shell, leading to much of the feasible region being reachable by these two simple geometries.  We also observe that the homogeneous spheres nearly reach the individual upper bound on scattering.  The same is not true for the larger electrical size ($ka = 1$, right panel), where the response of shells lies within the ``interior'' of the locus traced by homogeneous spheres of varying radii.  Note that in both cases, the maximum achievable scattering cross section is orders of magnitude larger than the maximum absorption cross section, consistent with the results seen in Fig.~\ref{fig:dielectric-individual}.

In order to further explore the feasibility of the derived bounds, shape optimization (i.e., inverse design) was performed within a spherical design region of radius~$a$ using a gradient-based technique~\cite{Capeketal_ShapeSynthesisBasedOnTopologySensitivity}, iteratively updated by the NSGA-II genetic algorithm~\cite{Deb_MultiOOusingEA}. Due to the high computational complexity of inverse design over a large number of design variables ($4173$~degrees of freedom), only maximal extinction cross section~$\sigma_\mathrm{t}$ ($w_\T{a} = w_\T{s}$) was optimized, see the diamond mark in Fig.~\ref{fig:dielectric-feasible}, right.  From this result, we can conclude that it may be possible to exceed the performance of simple structures through the use of automated synthesis methods, however, any improvement must be limited by the fundamental bounds defining the feasible objective space. In addition, we hypothesize that the performance of inverse designed structures may be improved by refining the initial discretization (increasing level of attainable details). The role of this effect is a subject of ongoing research.

\subsection{Example: metallic scatterers}
\label{sec:mat-au}

\begin{figure}
    \centering
    \includegraphics[width=3.5in]{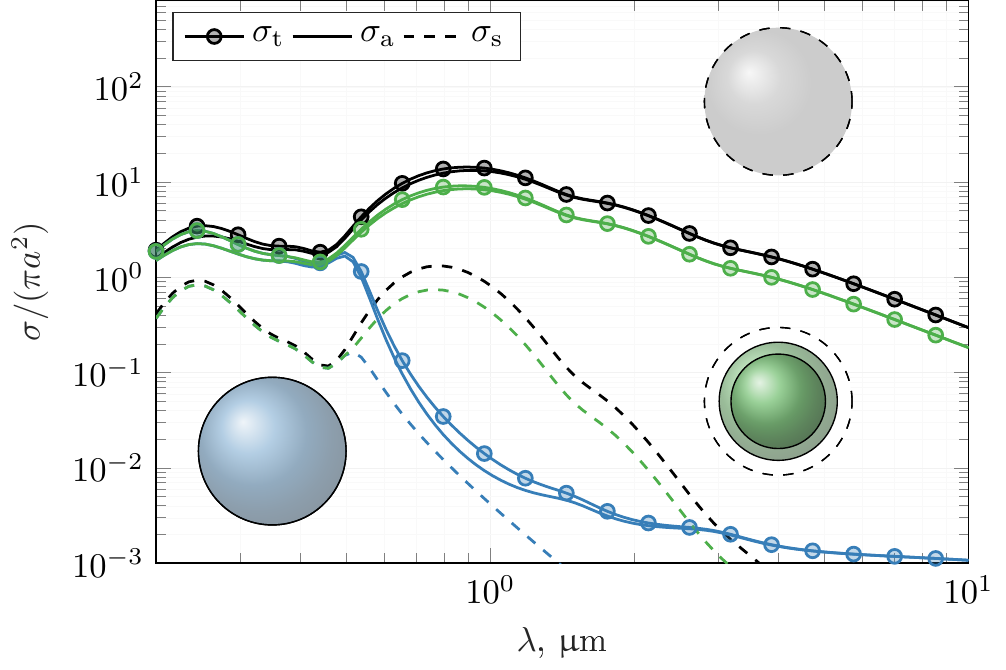}
    \caption{Maximal cross sections for a gold (Au) sphere of radius 30~nm (black) along with the realized cross sections of a homogeneous Au sphere (blue) and an optimized Au shell (green).}
    \label{fig:au-individual}
\end{figure}

As a second example, we consider optimal cross sections for devices constructed of gold (Au) confined to spherical regions of radius $a=30$~nm.  For these calculations, the local, frequency dependent Lorentz-Drude permittivity model \cite{rakic1998optical,ung2007interference} is used.  Single-objective optimal cross sections are shown in Fig.~\ref{fig:au-individual} along with the realized cross sections from solid spheres of radius $r=a$ and spherical shells with inner and outer radii optimized for maximum absorption, scattering, and extinction.  

Unlike in the previous example, homogeneous Au spheres of radius $r<a$ are far from the single objective bounds in the low frequency regime.  However, optimizing a spherical shell by sweeping over both inner and outer shell radii closely approaches the single objective bounds (within one order of magnitude) over all frequencies, suggesting that this simple geometry is capable of nearly reaching optimal performance given the material and geometric design constraints.

\begin{figure}
    \centering
    \includegraphics[width=2.25in,height=2.25in]{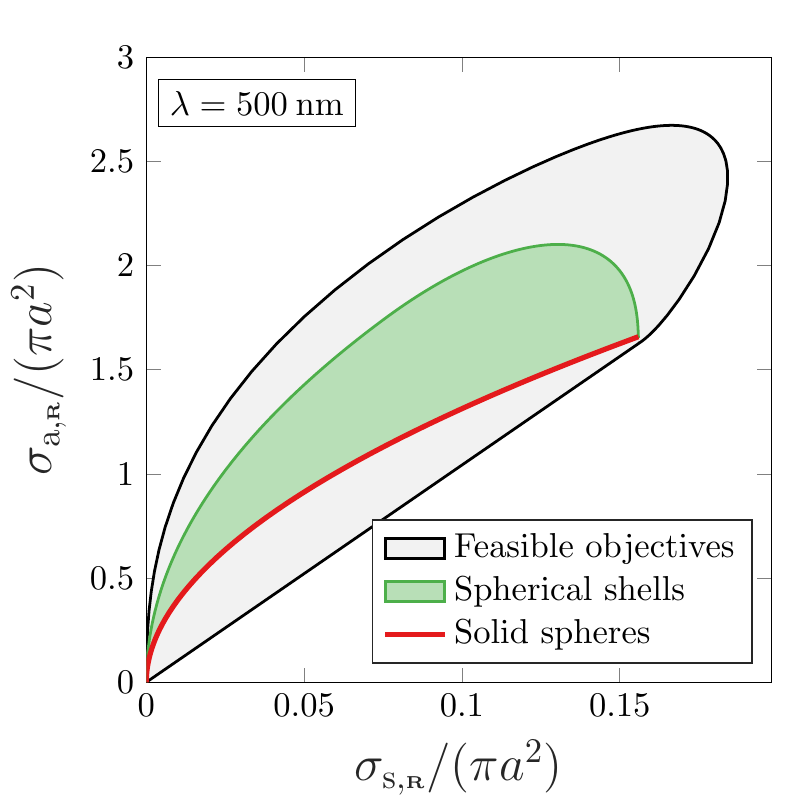}
    \includegraphics[width=2.25in,height=2.25in]{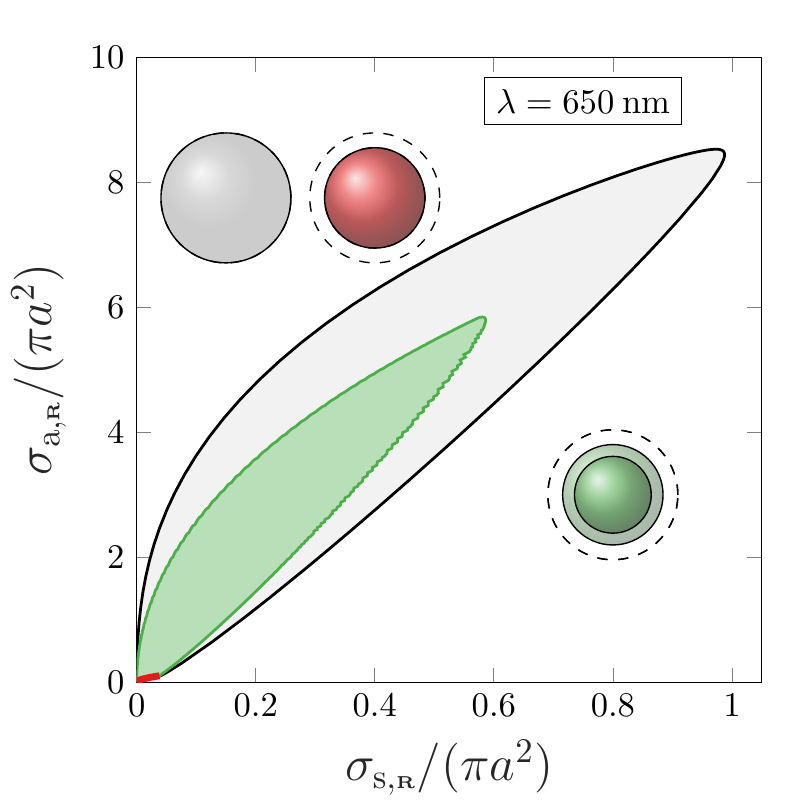}    
    \caption{Feasible region for Au obstacles fitting within a sphere of radius a $a=30\unit{nm}$ and realized cross sections of solid spheres (red lines) and spherical shells (green region).}
    \label{fig:Au-realized}
\end{figure}

In Fig.~\ref{fig:Au-realized}, we compare the feasible objective space for this configuration at two wavelengths, $\lambda = 500~\T{nm}$ and $\lambda = 650~\T{nm}$, against the realized behavior of solid spheres with varying radius $r\leq a$ (red color) and air-core Au shells with maximum outer radius $r \leq a$ (green color).  At $\lambda = 500~\T{nm}$, we observe that the solid spheres and spherical shells cover comparable portions of the feasible objective space, both reaching nearly the maximum value of absorption and scattering cross sections.  At $\lambda = 650~\T{nm}$, the situation is quite different, with solid spheres covering very little of the feasible objective space compared to spherical shells.  This indicates that, in scenarios such as this one, there are opportunities to significantly enhance absorption and / or scattering through the use of geometries more complex than a solid sphere.

\section{Conclusions}
\label{sec:conclusions}

The bounds derived in Secs.~\ref{sec:losses} and~\ref{sec:materials} are general for any substructures contained with arbitrarily shaped design regions.  The only restriction is that the substructure is formed by replacing parts of the (possibly inhomogeneous) resistivity of the design region with that of the background medium.    Generalizations toward mixed-material substructures is the subject of ongoing and future work.  In this paper, we focus on spherically-symmetric design regions, though this need not be the case.  Rather, any region describable by an impedance matrix and its components (cf Sec.~\ref{sec:MOO}) may be studied.  Generating these matrices for arbitrary geometries are straightforward using the method of moments, either by local volumetric discretization \cite{polimeridis2014stable} or whole domain basis functions \cite{mautz1969radiation}.

The distinction between the prescribed losses and materials problems in Secs.~\ref{sec:losses} and \ref{sec:materials} warrants close examination on several points.  The prescribed materials case, where the complete complex resitivity of the design material is specified, clearly illustrates the physical limitations of single-material nanophotonic structures, e.g., nano-antennas constructed of gold confined to a certain design volume.  The multiobjective bounds derived here not only indicate the feasible performance property of any realized structure, they also indicate that, in specific cases, near-optimal performance may be obtained by extremely simple geometric paramterizations.  In those cases, such as the metallic scatterers examined in Sec.~\ref{sec:mat-au}, we conclude that there is little performance gain to be had through computationally expensive inverse design.  The exact quantification of further gains must be made on a case-by-case basis, but the comparison between the feasible objective space and the performance of simple parameterized spherical structures is computationally inexpensive, requiring only the solution of the appropriate multiobjective optimization problem and a swept parameterization of simple test structures.  If the test structures are chosen to be spherically symmetric, the computational overhead of computing their scattering and absorption performance via Mie theory is negligible \cite{bohren2008absorption}.  
While less specific than the prescribed materials bounds, the prescribed losses bounds represent limits on future materials that may have fixed losses but tunable reactances.  Tunable reactance may be possible through the use of metamaterials (e.g., subwavelength patterning of two or more materials).  The effect of the fixed amount of loss within such a material is quantified directly by the bounds in Sec.~\ref{sec:materials}, which may serve as a figure of merit in future nanophotonics material development.  Notably, these limits on feasible performance are even less computationally expensive than the prescribed materials bounds and, in several cases, admit analytic solutions, see Secs.~\ref{sec:losses-minmax} and \ref{sec:losses-esa}.  As in the prescribed materials case, the assessment of parameterized simple geometries is straightforward, allowing for fast comparison of the fundamental bounds to the performance of simple test structures, as demonstrated in Secs.~\ref{sec:losses-realized}.  The analysis of realized non-spherical structures with homogeneous resistivity can be greatly accelerated over conventional repeated solution of \eqref{eq:vzi}, allowing for their rapid evaluation in addition to Mie-based analysis of systems with spherical symmetry.

Throughout the entirety of this paper, we consider only isotropic, homogeneous, bulk, linear material properties.  The inclusion of anisotropic materials is straightforward by adapting the resistivity (or susceptibility) into a tensor form.  In this case, statements regarding the dependence of certain quantities on material parameters, e.g., the linear real resistivity dependence of the loss matrix $\M{R}_\rho$, must be slightly reinterpreted.  The same holds for studying inhomogeneous material properties within the design region. Inclusion of surface effects in nanoscale objects is also compatible with all formulations in this work, requiring only small adaptations to the calculation of the impedance matrix using a linear non-local hydrodynamic model \cite{mortensen2013nonlocal}, see \cite[\S D]{gustafsson2020upper}.

In conclusion, the formulations presented in this work allow for the determination of feasible absorption and scattering characteristics of objects confined to prescribed design regions and constructed with material properties of varying degrees of specification.  The example calculations demonstrate that the question \emph{``Are the optimal absorbers also optimal scatterers?''} is nuanced and depends on the exact nature of the system under consideration.  Nonetheless, the formulations in this work provide bounds that can inform and direct future work in the inverse design of optimal photonic devices.  Additionally, the multiobjective framework presented here may be adapted to study trade-offs between absorption, scattering, and other performance objectives, such as near-field enhancement or Purcell factor.

\section*{Funding}
Grantov\'{a} Agentura \v{C}esk\'{e} Republiky (19-06049S); Vetenskapsr\aa det (2017-04656); Henry Luce Foundation.

\ifosa
\else
\bibliographystyle{IEEEtran}
\bibliography{refs}
\fi

\end{document}